\documentclass{article}
\pdfoutput=1
\usepackage[letterpaper, portrait, margin=1.5in]{geometry}
\usepackage[T1]{fontenc}
\usepackage{color}
\usepackage{nicefrac}
\usepackage{multirow}
\usepackage{float}
\usepackage[english]{babel}
\usepackage{amsmath, amssymb, mathtools}
\usepackage{fancyhdr}
\usepackage{scrextend}
\usepackage{csquotes}
\title{\textbf{Anomalous Postselection\\in the Hensen et al. Bell Test}}
\author{Donald A. Graft\\\textit{donald.graft@cantab.net}}
\date{}

\begin{document}
\lfoot{}
\cfoot{}
\rfoot{}
\lhead{}
\chead{}
\rhead{}
\maketitle
\thispagestyle{empty}

\begin{abstract}
It is shown that the data of the Hensen et al. Bell test experiment exhibits anomalous postselection that can fully account for the apparent violation of the CHSH inequality. A simulation of a local realist model implementing similar postselection is presented. The model produces an apparent violation of CHSH indistinguishable from that of the experiment. The experimental data also appears to violate no-signaling, and it is shown how postselection can produce an artifactual violation of no-signaling. The Hensen et al. experiment does not succeed in rejecting classical locality and therefore does not confirm quantum nonlocality.
\end{abstract}

\rhead{\thepage}

\section{Introduction}
\label{Introduction}

Interpreted in modern terms (rather than in terms of the debate over completeness of quantum mechanics), the 1935 paper of Einstein, Podolsky, and Rosen \cite{Einstein00}, through its demonstration that the orthodox rule for quantum state projection (L{\"u}ders' rule) implies superluminal transmission of influences, opened a furious debate in the foundations of physics that rages to this day. John Bell \cite{Bell00} made a seminal contribution to the debate by developing inequalities combining the results of correlations for event detections in scenarios with different combinations of measurement settings. These inequalities transformed what was previously essentially a philosophical debate into an empirical matter, decidable by experiment, and they have subsequently been used to attempt to distinguish between a fully local universe and one that allows for nonlocal interactions. These nonlocal interactions have come to be referred to as `quantum nonlocality'. 

Many modern experiments, including that of Hensen et al. \cite{Hensen00,Hensen01}, use a particular Bell-type inequality, the CHSH inequality \cite{CHSH}. These experiments are widely believed to confirm the existence of quantum nonlocality. However, these experiments have been subject to `loopholes', purported to allow local classical mechanisms to reproduce the inequality violation reported in the experiments. It has been a longstanding quest of experimental physicists to close all the loopholes and thereby decisively confirm the existence of quantum nonlocality. The experiment of Hensen et al. is the most prominent and arguably most convincing of the recent experiments claimed to be loophole-free. If the experiment was to be accepted as fully valid and decisive, the foundations of physics would be rocked.

Despite the claim by Hensen et al. that their experiment is loophole-free, at least one important loophole, the postselection loophole, remains open. The postselection loophole requires (in the general case), through a mechanism not specified, that some fraction of the full data of the experiment is absent from or not considered in the data analysis, producing only an artifactual violation of the applicable Bell-like inequality. Experimental physicists conducting an experiment must prove that postselection is not present in the experiment, or that any postselection that is present is harmless. Here I discuss the postselection loophole in the context of the Hensen et al. experiment, and I argue that Hensen et al. have not succeeded in discharging their responsibility to prove the absence of harmful postselection.

Throughout the paper, the term `postselected' means postselected {\it out}, i.e., decimated. In some other contexts, the term might be used to denote selective inclusion, but here the meaning is selective exclusion.

The scheme of the paper is as follows. Section 2 demonstrates that postselection occurs in the Hensen et al. experiment. Section 3 shows how similar postselection for a classical local model can produce results indistinguishable from the results of the experiment, including a significant violation of the CHSH inequality. Section 4 shows how apparent violation of no-signaling in the experimental data further strengthens the arguments of this paper. Section 5 discusses these results and concludes that the Hensen et al. experiment does not succeed in rejecting local realism, and that a local realist may take it as confirming locality.

\section{Demonstration of anomalous postselection in the data of the Hensen et al. Bell test}
\label{Demonstration of anomalous postselection in the data of the Hensen et al. Bell test}

To investigate the matter of postselection in the Hensen et al. experiment, I instrumented the publicly available Hensen et al. analysis code to test the fairness and uniformity of the random number generators (RNGs) used in the experiment. Specifically, I replaced the appropriate lines in the Hensen et al. analysis code with the following lines:

{\footnotesize \fontfamily{ccr}\selectfont \begin{verbatim}
random_number_A             = data[:,6].astype(bool, copy=False)
random_number_A_not         = ~random_number_A
random_number_B             = data[:,7].astype(bool, copy=False)
random_number_B_not         = ~random_number_B
random_00                   = random_number_A_not & random_number_B_not
random_10                   = random_number_A & random_number_B_not
random_01                   = random_number_A_not & random_number_B
random_11                   = random_number_A & random_number_B
print('number of random 0\'s at A =', np.sum(random_number_A_not))
print('number of random 1\'s at A =', np.sum(random_number_A))
print('number of random 0\'s at B =', np.sum(random_number_B_not))
print('number of random 1\'s at B =', np.sum(random_number_B))
print('number of random 00 events =', np.sum(random_00))
print('number of random 01 events =', np.sum(random_01))
print('number of random 10 events =', np.sum(random_10))
print('number of random 11 events =', np.sum(random_11))
\end{verbatim}}

This instrumentation allows one to assess the fairness of the list of random setting choices for each side individually and the distribution of the four joint setting combinations. When the instrumented analysis code is run on the data of the experiment, the following results are obtained (the published Hensen et al. results are, of course, reproduced):
\newpage
{\footnotesize \fontfamily{ccr}\selectfont \begin{verbatim}
number of random 0's for A = 2340
number of random 1's for A = 2406
number of random 0's for B = 2345
number of random 1's for B = 2401
number of random 00 events = 1143
number of random 01 events = 1197
number of random 10 events = 1202
number of random 11 events = 1204
----------------------------------------
k/n: 196.0/245
p-value : 0.039

xy     ++,+-,-+,--
ab 00 [23  3  4 23]   (0,    -3pi/4)
ab 01 [33 11  5 30]   (0,    +3pi/4)
ab 10 [22 10  6 24]   (pi/2, -3pi/4)
ab 11 [ 4 20 21  6]   (pi/2, +3pi/4)

 E (RND00  RND01  RND10  RND11 )
   (+0.736, +0.595, +0.484, -0.608) measured
+/-( 0.093,  0.090,  0.111,  0.111 )
CHSH S : 2.422 +- 0.204
\end{verbatim}}

The additional code instrumentation shows a large excess of 1's at both A and B,  and a large deficit of events for the \{00\} experiment. Let us now estimate the probability that the joint counts distribution we see in the experiment could be obtained by chance. We see that the \{00\} experiment has a low count at 1143 while the other three experiments all have counts that exceed the expected mean value of 1186. I wrote and executed a numerical simulation that estimates the probability that the \{00\} count is less than or equal to 1143. The resulting p-value is 0.07. While this result is not significant at an arbitrary  5\% level, it shows that it is unlikely that the observed distribution could be produced by chance. The code for the numerical simulation is available on-line \cite{code00}.

My argument here does not depend on an exact p-value, but rather on a demonstration that it is unlikely that the observed distribution could be obtained by accident, together with  consideration of a local model violating CHSH using postselection and the presence of a no-signaling violation in the experimental data (see Section 4). Taken together, I argue that the 
Hensen et al. data for the \{00\} experiment was postselected.

The look-elsewhere effect is not included because I test the specific null hypothesis that the \{00\} experiment is postselected (discussed further in Section 5). This hypothesis was chosen in advance because the \{00\} experiment is special for CHSH. It is the `converse' to the \{11\} experiment, which is the only term in CHSH that is negatively weighted, and so if one wanted to choose a term to postselect so as to have a maximum effect on the CHSH metric, and so as to minimize the amount of required postselection, one would choose the \{00\} experiment. The charge of `harking' can therefore be rejected.

\section{Effect of anomalous postselection on Bell test results}
\label{Effect of anomalous postselection on Bell test results}

Acting as Devil's Advocates (see Section 5), let us suppose that the \{00\} postselection loses only mismatch events (events with opposite outcomes). The failure of this supposition does not weaken the conclusions of Section 2. To investigate the effect of this postselection I created a local numerical simulation model saturating the CHSH inequality at the classical limit of 2, together with variable postselection of \{00\} mismatch events (the number of lost events can be selected via user input when running the simulation). The number of events per run is set to 245 as for the Hensen et al. experiment (there are no entanglement failure events in a simulation), and 100000 runs are performed to obtain estimates for the mean CHSH metrics $S$ and $k$, and the mean counts characterizing the distribution of settings. For example, I give below the result of running the simulation with 14 lost \{00\} mismatch events. It can be seen that the postselection artifactually increases $S$ and $k$ beyond the classical limits. The closeness of the simulated $S$ and $k$ values to the Hensen et al. reported values of $S = 2.42$ and $k = 196$, as well as the similarity of the patterns of counts characterizing the distribution of settings, are stunning.

{\footnotesize \fontfamily{ccr}\selectfont \begin{verbatim}
mean number of 0's for A = 109
mean number of 1's for A = 122
mean number of 0's for B = 109
mean number of 1's for B = 122
mean number of 00 events = 47
mean number of 01 events = 61
mean number of 10 events = 61
mean number of 11 events = 61
----------------------------------------------
mean S = 2.414843
mean k = 197.250380
\end{verbatim}}

Figure 1 shows the relationship between the number of lost events and $S$. Loss of 14 or more events will reproduce or exceed the reported Hensen et al. results. The code for the numerical simulation model is available on-line \cite{code01}.\\
\\
\centerline {Figure 1. Effect of postselected \{00\} mismatch events on S}
\centerline {for a local realist model operating at the classical limit}

\begin{center}\includegraphics[scale=0.4]{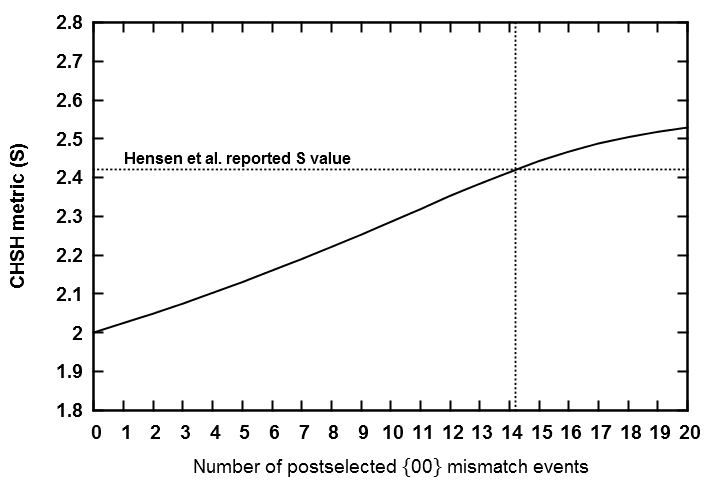}\end{center}

\section{Violation of no-signaling provides further evidence of postselection}
\label{Violation of no-signaling provides further evidence of postselection}

Bednorz \cite{Bednorz} and later Adenier and Khrennikov \cite{Adenier} demonstrated statistically significant violation of the no-signaling criterion in the data of the Hensen et al. experiment. Postselection of \{00\} events produces an asymmetry that results in an artifactual violation of no-signaling. It is not difficult to see how this works. Consider side A's counts of 1's for the \{00\} and \{01\} experiments. If side B's setting is not affecting the counts, as required by no-signaling, then these two side A counts should be close to each other. However, postselection of \{00\} events will selectively reduce the first side A count, thereby biasing the results and producing an artifactual violation of no-signaling. This is confirmed by numerical simulation.

The finding of apparent no-signaling violation in the Hensen et al. data therefore strengthens the analysis of this paper. Note that neither Bednorz nor Adenier and Khrennikov discussed postselection, and they did not discover the anomalous postselection in the Hensen et al. data that I report here.

\section{Discussion}
\label{Discussion}

In an interesting thread at {\it PubPeer} \cite{PubPeer} discussing the Hensen et al. experiment, `Peer 6' notices the missing events in the public release of data by Hensen et al. With my own grammatical corrections and paraphrasing I give the main points made by `Peer 6':

\begin{displayquote}
1) The published data does not contain raw data. Instead, it contains preprocessed data, because the raw data from stations A, B, and C have already been brought together into a single file, and the file contains only 4746 events, whereas the paper reports orders of magnitude more events.
  
2) In the supplementary information \cite{Hensen01}, Hensen et al. say: ``Every few hundred milliseconds, the recorded events are transferred to the PC. During the experiment, about 2 megabytes of data is generated every second. To keep the size of the generated data-set manageable, blocks of about 100000 events are saved to the hard drive only if an entanglement heralding event is present in that block.'' Therefore, what is published (4746 events, approximately 420 kilobytes) is only about 5\% of a single block.
\end{displayquote}

While elimination of blocks without valid events is legitimate (although it reduces the precision of possible RNG tests in the analysis), the deletion process also allows for deletion of \{00\} mismatch events. The processing to transform the individual raw lists to the published combined list is neither documented nor published, and the individual raw lists themselves are not available for inspection. This processing is exactly the critical step where postselection can occur. The withholding of the data and processing is troubling, given the importance of the result for the foundations of physics.

Hensen et al. address the distribution of randomness for setting choices in the experimental data, arguing that it is fair and uniform \cite{Hensen02}. However, the counts they provide are insufficient to expose the anomaly reported here. Hensen et al. write:

\begin{displayquote}
``We can get further insight by looking at all the setting choices recorded during the test. Around every potential heralding event about 5000 settings are recorded, for which we find a local P-value of 0.57 (Table 1), consistent with a uniform setting distribution.''
\end{displayquote}

Unfortunately, the analysis code and data justifying this conclusion are not available and the claim of uniformity appears to be false for the published data, as I demonstrated in Section 2. The overall counts in Table 1 apparently contain all the excluded blocks but the CHSH calculation is performed only over the published data. If the overall counts are indeed highly uniform as claimed while the published subset of the overall data is not uniform, the conclusion that the published data was postselected is further strengthened. One also wonders why great pains are taken by Hensen et al. to show uniformity for the overall data, but nothing is said about the uniformity of the published subset of the data.

Understandably, one might be expected to speculate about the specific mechanism of the postselection demonstrated here. Given the transcending importance of the result for the foundations of physics, including Einstein's legacy therein, it is appropriate to address matters from two stances. The first stance is that of the Devil's Advocate, a stance that dispenses with decorum and `speaks truth to power'. The second stance is that of the Dispassionate Observer, a stance that refrains from speculating on matters not in full evidence.

The Devil's Advocate reasons as follows. Consider the null hypothesis that Hensen et al. did not manually (possibly inadvertently) postselect \{00\} mismatch events. Is this hypothesis rejected by the evidence of this paper? Given that the probability of obtaining the observed distribution of random joint setting choices in the experimental data by chance is 0.07 (Section 2), that an apparent violation of no-signaling is observed in the data, and that a local model that postselects \{00\} mismatch events produces results indistinguishable from the experimental results, the null hypothesis can be rejected, i.e., Hensen et al. manually (possibly inadvertently) postselected \{00\} mismatch events.

The Dispassionate Observer reasons as follows: Consider the null hypothesis that Hensen et al. did not manually (possibly inadvertently) postselect \{00\} events. Is this hypothesis rejected by the evidence of this paper? Given that the probability of obtaining the observed distribution of random joint setting choices in the experimental data by chance is 0.07 (Section 2), and that an apparent violation of no-signaling is observed in the data, the null hypothesis can be rejected, i.e., Hensen et al. manually (possibly inadvertently) postselected \{00\} events.

The difference between the two stances is subtle. The Devil's Advocate asserts in the null hypothesis that the postselected events are \{00\} mismatch events, while the Dispassionate Observer does not (however, both conclude that the \{00\} data is postselected). While the reasoning of the Devil's Advocate already appears compelling, a direct determination between the stances could easily be made by inspecting the individual raw lists. However,  these raw lists have not been released (Hensen et al. have thus far declined to provide access to them). Nevertheless, regardless of which stance we adopt, the experiment must be considered to be placed in doubt.

It is theoretically possible, of course, that defective operation of the devices of the experiment, or improper design of the experiment could account for the observed postselection. However, it is very difficult to conceive of such defects, and so this possibility can be considered to be implausible. Nevertheless, if this were the case, the experiment would again clearly be placed in doubt.

The design and implementation of the Hensen et al. experiment is laudable, and experiments like it offer the prospect of deciding the debate over nonlocality. However, the experiments can be decisive only when the data and analyses are correct and transparent, and when undocumented steps are not present in the analysis and interpretation.

One could consider analyzing the second run of the Hensen et al. experiment to see if similar postselection appears. However, Hensen et al. concede that an equipment failure occurred in the middle of the run, placing the previously recorded data in doubt. I choose therefore, for the purposes of this paper, to confine the discussion to the first, successfully executed run.

My personal views on quantum nonlocality \cite{Graft00,Graft01,Graft02,Graft03} are by now hopefully well-known. I argue that the correct quantum prediction for EPR must not use L{\"u}ders' rule, which means that nonlocal correlations are not predicted by quantum mechanics, and that the experiments, when properly designed, analyzed, and interpreted,
confirm locality and disconfirm quantum nonlocality. Einstein's legacy and Lorenz invariance are safe, and physics remains consistent and coherent.

\renewcommand\refname{References}


\begin{thebibliography}{99}

\bibitem{Einstein00} A. Einstein, B. Podolsky, and N. Rosen, ``Can Quantum-Mechanical Description of Physical Reality Be Considered Complete?" {\it Physical Review},
 {\bf 47}, 777-780 (1935).

\bibitem{Bell00} J. S. Bell, {\it Speakable and Unspeakable in Quantum Mechanics},
2nd ed., Cambridge University Press, Cambridge (1987).

\bibitem{Hensen00} B. Hensen, H. Bernien, A. E. Dr\'{e}au, A. Reiserer, N. Kalb, M. S. Blok, 	J. Ruitenberg, R. F. L. Vermeulen, R. N. Schouten, C. Abell\'{a}n, W. Amaya, V. Pruneri, M. W. Mitchell, M. Markham, D. J. Twitchen, D. Elkouss, S. Wehner, T. H. Taminiau, and R. Hanson, ``Loophole-free Bell inequality violation using electron spins separated by 1.3 kilometres'', {\it Nature} {\bf 526}, 682-686 (2015).

\bibitem{Hensen01} B. Hensen, H. Bernien, A. E. Dr\'{e}au, A. Reiserer, N. Kalb, M. S. Blok, 	J. Ruitenberg, R. F. L. Vermeulen, R. N. Schouten, C. Abell\'{a}n, W. Amaya, V. Pruneri, M. W. Mitchell, M. Markham, D. J. Twitchen, D. Elkouss, S. Wehner, T. H. Taminiau, and R. Hanson, ``Loophole-free Bell inequality violation using electron spins separated by 1.3 kilometres'', {\it Nature} {\bf 526}, Supplementary Information 15759 (2015).

\bibitem{CHSH} J. F. Clauser, M. A. Horne, A. Shimony, and R. A. Holt, ``PROPOSED EXPERIMENT TO TEST LOCAL HIDDEN-VARIABLE THEORIES'', {\it Phys. Rev. Lett.} {\bf 23}, 880-884 (1969).

\bibitem{code00} \begin{flushleft}
The simulation code is available at the following URL: \small \fontfamily{ccr}\selectfont http://rationalqm.us/papers/Hensen/calc.cpp\end{flushleft} \normalfont \normalsize

\bibitem{code01} \begin{flushleft}
The simulation code is available at the following URL: \small \fontfamily{ccr}\selectfont http://rationalqm.us/papers/Hensen/chsh.cpp\end{flushleft} \normalfont \normalsize

\bibitem{Hensen02} B. Hensen, N. Kalb, M. S. Blok, A. E. Dr\'{e}au, A. Reiserer, R. F. L. Vermeulen, R. N. Schouten, M. Markham, D. J. Twitchen, K. Goodenough, D. Elkouss, S. Wehner, T. H. Taminiau, and R. Hanson, ``Loophole-free Bell test using electron spins in diamond: second experiment and additional analysis'', arXiv:1603.05705 [quant-ph] (2016).

\bibitem{Bednorz} A. Bednorz, ``Signaling loophole in experimental Bell tests'', arXiv:1511.03509 [quant-ph] (2015).

\bibitem{Adenier} G. Adenier and A. Khrennikov, ``Test of the no-signaling principle in the Hensen loophole-free CHSH experiment'', arXiv:1606.00784 [quant-ph] (2016).

\bibitem{PubPeer} \begin{flushleft}
Discussion thread on PubPeer with contributions by `Peer 6': \small \fontfamily{ccr}\selectfont https://pubpeer.com/publications/B74AEB0DD4EF02CB25D6F39C2CFE88\end{flushleft} \normalfont \normalsize

\bibitem{Graft00} D. A. Graft, ``On reconciling quantum mechanics and local realism", Proceedings of SPIE conference ``The Nature of Light: What are Photons? 5", SPIE, Bellingham (2013). Also arXiv: quant-ph 1309.1153 (2013).

\bibitem{Graft01} D. A. Graft, ``Analysis of the Christensen et al. Test of Local Realism'', {\it J. Adv. Phys.} {\bf 4}(3), 284-300 (2015).

\bibitem{Graft02} D. A. Graft, ``Clauser-Horne/Eberhard inequality violation by a local model'', {\it Adv. Sci. Eng. Med.} {\bf 8}, 496-502 (2016).

\bibitem{Graft03} D. A. Graft, ``The Quantum Prediction for Einstein-Podolsky-Rosen (EPR) Experiments'', arXiv:1607.01808 [quant-ph] (2016).

\end{thebibliography}
\end{document}